\documentclass[10pt,letterpaper]{article}
\usepackage[top=0.85in,left=2.75in,footskip=0.75in]{geometry}

\usepackage{amsmath,amssymb}

\usepackage{changepage}

\usepackage[utf8x]{inputenc}

\usepackage{textcomp,marvosym}

\usepackage{cite}

\usepackage{nameref,hyperref}

\usepackage[right]{lineno}

\usepackage{microtype}
\DisableLigatures[f]{encoding = *, family = * }

\usepackage[table]{xcolor}

\usepackage{array}

\newcolumntype{+}{!{\vrule width 2pt}}

\newlength\savedwidth

\raggedright
\usepackage[aboveskip=1pt,labelfont=bf,labelsep=period,justification=raggedright,singlelinecheck=off]{caption}

\makeatletter
\renewcommand{\@biblabel}[1]{\quad#1.}
\makeatother

\usepackage{lastpage,fancyhdr,graphicx}
\usepackage{epstopdf}
\fancyheadoffset[L]{2.25in}
\fancyfootoffset[L]{2.25in}
\DeclareMathOperator{\dist}{dist}

\DeclareMathOperator{\vol}{vol}
\usepackage{epstopdf}
\AppendGraphicsExtensions{.tif}

\begin{document}
\vspace*{0.2in}

\begin{flushleft}
{\Large
\textbf\newline{On node ranking in graphs} % Please use "sentence case" for title and headings (capitalize only the first word in a title (or heading), the first word in a subtitle (or subheading), and any proper nouns).
}
\newline
\\
Ekaterina Dudkina,\textsuperscript{1}
Michelangelo Bin,\textsuperscript{2*}
Jane Breen,\textsuperscript{3}
Emanuele Crisostomi,\textsuperscript{1}
Pietro Ferraro,\textsuperscript{4}
Steve Kirkland,\textsuperscript{5}
Jakub Mare\u{c}ek,\textsuperscript{6} 
Roderick Murray-Smith,\textsuperscript{7}
Thomas Parisini,\textsuperscript{2}$^,$\textsuperscript{8}$^,$\textsuperscript{9}
Lewi Stone,\textsuperscript{10}$^,$\textsuperscript{11}
Serife Yilmaz,\textsuperscript{4}
Robert Shorten\textsuperscript{4} 
\\
\bigskip
\textbf{{1} Department of Energy, Systems, Territory and
Constructions Engineering, University of Pisa, Pisa, Italy.} \\
\textbf{{2} Department of Electrical and Electronic Engineering, Imperial College London, London, UK.}  \\
\textbf{{3} Ontario Tech University, Canada.} \\
\textbf{{4} Dyson School of Design Engineering, Imperial College London, London, UK.} \\
\textbf{{5} Department of Mathematics, University of Manitoba, Canada.} \\
\textbf{{6} Czech Technical University, Prague, Czech Republic.} \\
\textbf{{7} School of Computing Science, University of
Glasgow, Glasgow, Scotland.} \\
\textbf{{8} Department of Engineering and Architecture, University of Trieste, Trieste, Italy.}  \\
\textbf{{9} KIOS Research and Innovation Center of Excellence, University of Cyprus, Nicosia, Cyprus.}  \\
\textbf{{10} The George S. Wise Faculty of Life Sciences, Tel Aviv University, Israel.}  \\
\textbf{{11} Mathematics, School of Science, RMIT University, Melbourne, Australia.}  \\

\bigskip

\end{flushleft}

\section*{Abstract}
The ranking of nodes in a network according to their ``importance'' is a classic problem that has attracted the interest of different scientific communities in the last decades. The current COVID-19 pandemic has recently rejuvenated the interest in this problem, as it is related to the selection of which individuals should be tested in a population of asymptomatic individuals, or which individuals should be vaccinated first. Motivated by the COVID-19 spreading dynamics, in this paper we review the most popular methods for node ranking in undirected unweighted graphs, and compare their performance in a benchmark realistic network, that takes into account the community-based structure of society. Also, we generalize a classic benchmark network originally proposed by Newman for ranking nodes in unweighted graphs, to show how ranks change in the weighted case.

%\linenumbers

\section*{Introduction}

\subsection*{Motivation}

The recent outbreak of COVID-19 and the various attempts to find effective non-pharmaceutical 
interventions to mitigate the impact of this virus have highlighted both the potential value of network science, 
and also the pressing need to further understand complex interaction networks to support the development of efficient {\em machine learning} techniques for graphs. Several diverse scientific communities have already developed different techniques to better analyze and understand complex networks of interactions, and some examples are listed below.\newline 
\begin{itemize}
\item In the study of electrical grids, a graph with its vertices (system components) and links (interactions
or dependencies among components) can, for example, represent a power
system. Analyses of this grid using graph theory can assess the vulnerability of grid components
during cascade failure (a model called extended betweenness combines
network structure with electrical characteristics of the power grid \cite{yan_integrated_2014})
or help to integrate renewable energy sources (for example, networks with the small-world property might be more resistant to small variations in load and
generation \cite{Koeth_graphs_nodate}).\newline

\item In neuroscience, networks may represent, for example, connections between neurons and help to understand the architecture and development of the brain. In this context, detection of communities and bridges, help
to ``separate functionally related neural elements'' and to study
the flow of neural signals and information \cite{sporns_graph_2018}.\newline

\item Another common application of {\em graph theory} arises in the context of 
social networks where several applications have 
been developed, including that of understanding the propagation of epidemics over graphs \cite{hufnagel_forecast_2004}.\\
 \end{itemize}
 
There are many other research areas in which the application of graph theory has played an important role. However, despite such a great interest from different communities, COVID-19 has illustrated both the importance of graph theory and its enormous potential, but also its limitations in the eyes of policy makers. For example, as we write this note, there are more than 7000 results for the combination of ``COVID-19'' and ``graph theory'' on Google Scholar. Yet few of these results,
to the best of our knowledge, have actually influenced policy in response to COVID-19.  
This is despite the fact that all networks arising in studies prior to the pandemic share many
important features with networks in which disease propagates. All are ``driven by 
common organizing principles'' and obey fundamental laws \cite{barabasi_network_2016}, and 
in fact common mathematical methods can be applied to these. A natural question arises 
in this context {\em why is this the case}?\newline 

Answering the aforementioned question is not trivial. One feature of COVID-19, that certainly 
limited the usefulness of graph theory, is the lack of proper debate on the tradeoff between
network utility (saving lives) and the right of individuals to {\em cast-iron} guarantees of privacy. 
Many (but not all) techniques for analyzing networks suffered in this context 
due to the fact that they rely on a centralized knowledge of community structure, and thus, of personal acquaintances. Another important factor is the {\it fog} that surrounds this area. As we have mentioned, the field of \textit{network science} has benefited from the fact that it is applicable in many research areas. Consequently, overlapping  contributions have been made in diverse areas. This is both good and bad; good as progress has been  fast, but also bad as it makes parsing available results difficult, both from the perspective of understanding how they connect with each other, and from the perspective  of applying these results to \textit{real world} problems. Our objective in this paper is to try to partially address this latter issue. Motivated by our interest in the spread of COVID-19, and building  on our recent paper on this topic \cite{yilmaz_kemeny-based_2020}, the specific objective of
the present manuscript is to review the many existing indicators that have been proposed to rank nodes in graphs, and to compare them in terms of their effectiveness and efficiency (i.e., in terms of the
required computational effort for large-scale graphs) in the specific context of COVID-19. While we start with the classic case of undirected unweighted graphs, we also provide an interesting comparison for the weighted case as well, for which fewer results can be found in the literature.

\subsection*{Contributions}
This paper provides the following main contributions:
\begin{enumerate}
\item
First, we review the most popular centrality measures that have been used in the existing literature to rank nodes in networks. While many other review papers have been published on this topic (see \cite{gomez_centrality_2019,ronqui_analyzing_2015,oldham2019consistency}), in our review we also include some ranking methods which are not particularly popular, but have been recently proposed for epidemiology applications; for instance the Kemeny index \cite{yilmaz_kemeny-based_2020};
\item
Second, we propose a benchmark case study which can be used for comparison purposes. This includes a description of how to create a realistic network of contacts, which is taken from a work on epidemiological studies \cite{salathe_dynamics_2010}, and a simplified model to mimic the COVID-19 spreading dynamics;
\item
We compare many different indicators, taken from different scientific communities (e.g., epidemiology, operations research, graph theory, control theory, communications and computer science communities) and rank them on their ability to mitigate the spreading of the virus in the same network. Moreover, indicators are also compared in terms of their required computational burden. This is particularly important to predict their scalability in real-world large-scale social networks;
\item
Finally, we revisit the classic undirected unweighted network that had been proposed by Newman to support the need for new ranking algorithms (namely, Random Walk Betweenness \cite{newman_measure_2005}), and we extend it to the weighted case. Such an example allows us to observe what indicators may be more appropriate to use in the case of weighted networks (in the context of COVID-19, this may be useful if not all only contacts are registered, but also their duractions).
\end{enumerate}

\subsection*{Organization of the paper}
The paper is organized as follows: firstly, we review the state of the art and briefly describes the indicators that will be later considered in the comparisons. In the following section we present the benchmark case study of undirected and unweighted network and the indicators' comparison. Next, we describe the revisited network to compare indicators in the weighted framework, and show the corresponding obtained ranks of the nodes. Finally, we conclude our manuscript.

\section*{Review of the state of the art}
\label{Indicators}

In this section we review methods of ranking nodes by how `central' they are to the network. While many surveys of this type exist (see, for example \cite{gomez_centrality_2019,ronqui_analyzing_2015,oldham2019consistency}), the aim of the current work is to compare and contrast these and their effectiveness in the context of disease spread in a contact network. Furthermore, we consider alternative methods for ranking node centrality by considering how much each node's removal contributes to a change in some global measure of connectivity of the network. As such, we also review some network connectivity measures, and their interpretation in the context of disease spread.

\subsection*{Some mathematical preliminaries}

A simple undirected graph is denoted $\mathcal{G} = (\mathcal{V}, \mathcal{E})$, with vertex set $\mathcal{V} = \{1,\hdots,N\}$ and edge set $\mathcal{E}\subseteq \{\{i,j\} \mid i,j \in \mathcal{V}\}$. We say that $i$ and $j$ are \emph{adjacent}, and denote this as $i\sim j$, if there is an edge between $i$ and $j$ (i.e. $\{i,j\} \in \mathcal{E})$. If $i\sim j$, we say that $j$ is a \emph{neighbour} of $i$. The number of neighbours of $i$ is referred to as the \emph{degree} of vertex $i$, and denoted $\deg(i)$. The adjacency matrix of $\mathcal{G}$ is the matrix $A$ defined entrywise as:
\[a_{ij}=\left\{\begin{array}{cc} 1, &\text{ if } i\sim j;\\ 0, & \text{otherwise}.\end{array}\right.\]

A directed graph with vertex set $\mathcal{V}$ as above is one in which the edges are \emph{ordered pairs} of vertices $(i,j)$; i.e. $\mathcal{E} \subseteq \mathcal{V}\times \mathcal{V}$. The edge (or arc) $(i,j)$ is considered to represent a connection from $i$ to $j$, sometimes denoted $i \to j$, to indicate that the arc's initial vertex is $i$ and terminal vertex is $j$. The adjacency matrix is defined in the same way as above, though $a_{ij} = 1$ if and only if $(i,j)\in \mathcal{E}$, and $a_{ij}\neq a_{ji}$ in general. 

A weighted graph arises when each edge $\{i,j\}$ is assigned a weight $w_{ij}$. This is taken into account in the adjacency matrix by simply allowing the $(i,j)$ entry of the matrix to be the weight $w_{ij}$.

Directed and weighted graphs provide generalizations of the simple undirected graph which may be very useful in the context of disease spread in a contact network. Differing weights of edges may account for differing strengths of transmission between pairs of individuals, due to, for example, different circumstances of the interaction---length of time, distance, etc. Asymmetric values allow for transmission to be more likely in one direction than the other between two individuals, for reasons more pertinent to the individual; for example, different levels of susceptibility, preventative measures, waning immunity, age differences, etc.

Many centrality measures and connectivity measures are defined for simple undirected graphs, and while some definitions may allow extensions to the weighted or directed case, they may lose some aspect of their interpretation; some do not generalize at all. In what follows, we attempt to indicate for each metric listed whether or not they do generalize in this way. The ones which extend most naturally, we find, are those which are derived from random walks, and so we include some mathematical preliminaries pertaining to these here.

A random walk on a connected graph $G$ (and here we shall consider strongly connected graphs in the directed case) is a discrete-time stochastic process in which, at any given time, a ``random walker'' occupies one vertex of the graph, and in a subsequent time-step, moves to an adjacent vertex $j$ of his/her current vertex $i$, according to some transition probability $p_{ij}$. For a simple random walk on an undirected graph, the transition probability $p_{ij}$ is simply $\frac{1}{\deg(i)}$; that is, the random walker chooses his/her next position uniformly at random from among the neighbours of his/her current vertex. This process is a Markov chain whose state space is the vertex set of $G$, since the state in any time-step depends only on the state of the chain in the previous time-step. The probability transition matrix $P=[p_{ij}]$ for this Markov chain is easily determined from the adjacency matrix $A$ by normalizing the rows so that they sum to 1. In the case of weighted graphs, the transition matrix is determined exactly the same way from the weighted adjacency matrix; it follows similarly in the case of directed graphs, although one runs into trouble if there are any vertices with no outgoing edges and the random walk is no longer well-defined. 

For an ergodic Markov chain with states indexed $1,\ldots, n$ and $n\times n$ transition matrix $P$, the stationary distribution vector of $P$, denoted $w$, is a left eigenvector of $P$ corresponding to the eigenvalue 1, normalized so that the entries of $w$ sum to 1 and thus represents a probability distribution across the states. In particular, $w_i$ represents the long-term probability that the Markov chain occupies the $i^{th}$ state. Note that in the case of a simple random walk on a connected undirected graph, the stationary distribution vector entries are proportional to the vertex degrees; 
\[w_i = \frac{\deg(i)}{\sum_k \deg(k)}.\] 
For any two states $i$ and $j$, the \emph{mean first passage time from $i$ to $j$}, denoted $m_{ij}$, is the expected time it takes to reach the $j^{th}$ state, given that the chain starts in the $i^{th}$ state.

\subsection*{Node centrality measures}

\subsubsection*{Degree centrality}

The first and simplest proposal for ranking nodes in a network is by considering each node's degree (sometimes called valency), or the number of neighbours of each node. The degree of node $i$ can be easily computed via the $i^{th}$ row sum of the adjacency matrix: 
\begin{equation}
\deg(i)=\sum_{j=1}^{N} a_{ij}.
\end{equation}
The degree is a simple centrality measure for undirected networks, where it identifies the most connected nodes, in the sense that highly-ranked nodes under this metric will have the most neighbours. In the context of disease spread in a contact network, these highly-ranked nodes correspond to individuals with the most contacts; as such, a high-degree node infected with the disease has more opportunity to spread the disease to other individuals.

In the case of directed graphs, one has to distinguish between incoming and outgoing edges, defining the \emph{in-degree} and \emph{out-degree}, respectively, as 
\[\deg_-(i) = \sum_j a_{ij}\]
\[\deg_+(i) = \sum_k a_{ki}.\]
Given this dual definition, it is more difficult to consider degree as a measure of centrality in the directed case; some options are to consider the sum or the average of the two (as in \cite{gomez_centrality_2019}).

For weighted graphs, the degree of a vertex is easily extended by defining $\deg(i)$ as the sum of the weights of incident edges. This is sometimes referred to as the \emph{strength} of a vertex (see, for example \cite{barrat_architecture_2004}).

\subsubsection*{Closeness centrality}

The \emph{distance} between nodes $i$ and $j$, denoted $\dist(i,j)$, is defined as the minimum number of consecutive edges needed to move from node $i$ to $j$ or, equivalently, as the length of a shortest path between them. The \emph{closeness centrality} of a node $i$ is computed by taking the inverse of the average distance from $i$ to any other node:

\begin{equation}
CC(i)=\frac{N}{\sum_j{\dist(i,j)}}.
\end{equation}

The larger the value of $CC(i)$, the more central node $i$ is in the network, in the sense that it is, on average, close to many other nodes.

As with degree centrality, the definition of closeness centrality can be extended to directed networks, though a distinction must be made on whether the distances are computed from, or to, the reference node $i$, respectively. Note also that ``distances'' are not symmetric in directed networks. %\textcolor{red}{Find reference to CC with directed graphs.} 
For weighted graphs, one could consider the edge-weights as a ``cost'' to traversing the edge, and thus define shortest-distance between $u$ and $v$ as the minimum weight of any path from $u$ to $v$ (where the weight of a path is the sum of the weights of edges in the path). With such a definition for distance in hand, it is reasonable to generalize the closeness centrality for weighted graphs; see \cite{opsahl_node_2010} for some limited discussion.

In a disease spread context, ranking vertices in a contact network by their closeness centrality would, in theory, highlight individuals for whom there is (on average) low degree of separation between the individual and all other members of the community. If this individual were infected, then, it takes fewer secondary infections on average to infect others.

\subsubsection*{Betweenness centrality}

The \emph{betweenness centrality} of a node $i$ is computed in terms of how many shortest paths pass through that node \cite{freeman_set_1977, freeman_centrality_1978}. In particular, fix a source node $s$ and target node $t$ (distinct from $i$), and let $\sigma_{st}$ denote the total number of shortest paths from $s$ to $t$ (or geodesics, i.e. paths of length $\dist(s, t)$). Letting $\sigma_{st}(i)$ denote the number of those paths that include node $i$, we take the ratio of these and then average over all choices for $s, t \neq i$:

\begin{equation}
\begin{array}{lll}
BC(i) & = & \dfrac{1}{(N-1)(N-2)} \displaystyle \sum_{\substack{s,t=1\\ s, t\neq i}}^{N}\dfrac{\sigma_{st}(i)}{\sigma_{st}}\\
& & \\
& = & \displaystyle \sum_{\substack{s=1}}^{N} \left( \sum_{\substack{t=s+1} }^{N}\frac{\sigma_{st}(i)}{\sigma_{st}}\right).\end{array}
\end{equation}

Accordingly, the betweenness centrality of a node corresponds to the fraction of shortest paths that pass across that node, and this expression is valid for both directed and undirected networks. In principle, betweenness centrality is expected to rank highly the nodes that behave as bridges between clusters in the network \cite{gomez_centrality_2019}. In the context of disease spread, these highly-ranked nodes would correspond to individuals who bridge multiple communities.

We encounter similar difficulties with the extension of betweenness centrality to weighted and directed graphs as with closeness centrality. Some limited work exists; see for example \cite{wang_betweenness_2008}, that focuses on the betweenness centrality of an edge in a weighted network, rather than betweenness centrality of nodes, and \cite{white_betweenness_1994}, that discusses several possible extensions to directed graphs and their limitations.

\subsubsection*{PageRank centrality}

The PageRank algorithm computes a ranking for every web page based on the graph of the World Wide Web. PageRank has applications in search engines and traffic estimation \cite{ilprints422}. The PageRank algorithm is best understood via a random walk on the network. 

PageRank can be thought of as a model of a ``random surfer'' who starts on a random webpage and keeps clicking on links randomly, never hitting ``back''; that is, he/she takes a random walk on the World Wide Web, a directed graph in which $(i,j)$ is an edge if webpage $i$ has a hyperlink to webpage $j$. At any point, the surfer may get bored and ``teleport'' to a random page in the network.  The long-term probability that the random surfer occupies webpage $i$ in this stochastic process is its PageRank---this corresponds to the stationary distribution of the corresponding Markov chain. Note that the transition matrix for this Markov chain is 
\[P = \alpha D^{-1}A + (1-\alpha)\tfrac{1}{N}J,\]
where $D$ is the diagonal matrix of vertex out-degrees, $A$ is the adjacency matrix, and $J$ is the $N\times N$ matrix of all ones \cite{Langville}. The parameter $\alpha$ is referred to as the \emph{damping factor} and represents the probability at each step that the ``random surfer'' will get bored and request another random page \cite{brin_anatomy_1998}. This is usually, as a matter of convention, set to $0.85$. We note that in the case of the World Wide Web, the underlying directed graph is not strongly-connected; indeed, there may be many webpages with no outgoing hyperlinks. This causes immediate problems with the random walk being well-defined, but is usually fixed by replacing any zero row of the adjacency matrix with a row of all 1s. Thus when the random surfer ends up on a webpage with no outgoing links, he/she chooses a webpage uniformly at random in the next step. This allows computation of the PageRank vector in the setting that some nodes have outdegree 0.

The interpretation of this measure of centrality is interesting in the context of disease spread. In the case that we work with a simple undirected graph, if the damping factor is set to $\alpha=1$ (i.e. a simple random walk with no teleportation), then the PageRank centrality ranking corresponds exactly to the node degree centrality ranking. However, including a damping factor may allow for the possibility that a person contracts the disease not from another individual they have contact with, but rather by chance (e.g. touching a surface with traces of the virus), or by some mechanism not accounted for in the contact network model. PageRank centrality ranking of the nodes can essentially be thought of as strongly related to the node degree ranking in this context, with some relaxation that may actually reflect the features of the disease more accurately.

If one wishes to consider a weighted or directed graph, PageRank centrality is a measure which extends naturally along with its interpretability, simply by considering a random walk on the given weighted or directed graph.

\subsubsection*{Random walk betweenness centrality}

A measure of betweenness centrality based on random walks, called {\it random walk betweenness (RWB)} was introduced in \cite{newman_measure_2005}. The idea is to calculate the centrality of a given node $i$ by the proportion of random walks that pass through node $i$. This is very similar to betweenness centrality, but does not consider shortest paths. In particular, the author of \cite{newman_measure_2005} describes it as ``the expected net number of times a random walk passes through vertex $i$ on its way from a source $s$ to a target $t$, averaged over all $s$ and $t$''. This measure was shown to better rank the importance of nodes in graphs with existing communities, and to be less correlated with vertex degree in most networks \cite{newman_measure_2005}. Interestingly, the method for computing this measure is strongly dependent on considering the graph in question as an electrical network and considering current flow through a vertex. The author then proves that this is equivalent to the ``flow'' of a random walk; however, this means that this particular definition of random walk betweenness does not extend to directed or weighted graphs, and there is no literature which attempts this.

Random-walk betweenness centrality is obtained by averaging the current flow ($f_i^{(st)}$) through vertex $i$ over all possible origins ($s$) and destinations ($t$),
\begin{eqnarray}
RWB(i) = \frac{2}{N(N-1)} \sum_{\substack{s,t=1 \\ s\neq t}}^N{f_i^{(st)}}.
\label{RWalkBetwenness}
\end{eqnarray}
%where $f_i^{(st)}$ is current flow through vertex $i$ from node $s$ to node $t$ which passes through node $i$. 
See \cite{newman_measure_2005} for further information on how this quantity is calculated.

While it is not necessarily evident from the electrical network description above, this value does indeed compute the expected net number of times a random walk would pass through node $i$ before reaching a target $t$, given that it starts at a source $s$, averaged over all pairs $s, t$. This intuitively seems like exactly the metric we are looking for when considering ``pivotal'' individuals in a contact network in which a disease is spreading, and as we will see in the next section, this metric is one which is particularly effective in controlling the disease when used exclusively to determine testing protocols. It is also known that in networks with strong community structure, immunization interventions targeted at individuals bridging communities (e.g., using random walk betweenness) are more effective than those simply targeting highly connected individuals \cite{salathe_dynamics_2010}. 

Note that the definition that can be directly applied only to simple undirected graphs; there does not exist a generalization to weighted or directed graphs. 

\subsubsection*{Random walk centrality (RWC)}

Random walk centrality is introduced in \cite{noh_random_2004}, and is said to ``quantify how central a node $i$ is located regarding its potential to receive information which is randomly diffusing over the network''. Given a graph $G$, consider a random walk on the graph with transition matrix $P$. The stationary distribution vector for $P$ is denoted $w$, and the $k^{th}$ entry of $w$ is denoted $w_k$. The characteristic relaxation time of vertex $k$ is introduced as $ \tau_k \equiv \sum_{j=0}^{\infty}((P^j)_{k,k}-w_k).$ 
This quantity converges whenever the transition matrix $P$ is primitive, which is the case for a random walk on a connected non-bipartite graph. The \emph{random walk centrality} \cite{noh_random_2004} of vertex $k$ is then calculated as:
\[C_k \equiv \frac{w_k}{\tau_k}.\] %\textcolor{red}{interpretation? how to compute $\tau_k$ in practice?}

In \cite{kirkland_random_2016}, the author observes that the random walk centrality is the reciprocal of the measure of centrality known as the \emph{accessibility index}. For a given vertex $k$, the accessibility index of vertex $k$ is defined
\[\alpha_k = \sum_{j\neq k} w_jm_{jk},\]
where $m_{jk}$ is the mean first passage time from $j$ to $k$. This definition allows the interpretation of the accessibility index as the expected time to reach vertex $k$ for the first time, given that we start in any randomly-chosen initial state. It is shown in \cite{kirkland_random_2016} that $C_k = 1/\alpha_k.$ This is particularly useful as it admits a definition of this as a measure of centrality not just for simple undirected graphs (as is introduced in \cite{noh_random_2004}) but for any Markov chain, as a measure of state centrality. In particular, this is a useful metric for both weighted and directed graphs. Further work on estimating the accessibility index may be found in \cite{johnson_estimating_2019}.

In the disease spread context, this measure ranks highly the vertices which are ``easily accessed'' from other vertices of the graph. This is a useful way to consider how ``central'' an individual is in a community. However, in the context of the spread of disease, and in particular when considering testing protocols in order to control the disease, this may not be appropriate. Controlling the disease in our simulations means determining individuals who are most instrumental in the disease spreading through the whole graph, and while those individuals with high random walk centrality may have increased likelihood of being infected, this may not coincide with 
individuals who ought to be tested and isolated as quickly as possible. 

\subsection*{Network connectivity measures}

Many criticality measures exist which provide a single, numerical value describing the ``connectedness'' of a network in some way. Such measures can be easily extended to measure the criticality of a single node $i$ in a graph $\mathcal{G}$ by inferring the criticality of the $i^{th}$ node from the change in criticality of the network after the $i^{th}$ node is removed from the network. This is described as follows in \cite{fouss_algorithms_2016}: given some graph invariant $cr(G)$ measuring the connectedness of the graph $G$ in some manner, define
\begin{equation*}
  cr_i(\mathcal{G}) := cr(\mathcal{G})-cr(\mathcal{G} \setminus i).
\end{equation*}
We include several such measures of node centrality here for consideration in later simulations.

\subsubsection*{Effective graph resistance (Kirchhoff index)}

The effective graph resistance is interpreted as a ``robustness measure'' of a network \cite{wang_kemenys_2017}. To formulate the effective graph resistance (also known as the total graph resistance, or the Kirchhoff index of the graph), the (undirected and connected) graph $G$ is seen as an electrical circuit, where
an edge $\{i,j\}$ corresponds to a resistor of $r_{ij}$ Ohm, and the effective graph resistance is the sum of the effective resistances over all pairs of vertices. These effective resistances $R_{st}$ can be computed using the Laplacian matrix of the graph.

The Laplacian matrix of a graph is defined as the adjacency matrix subtracted from the diagonal matrix of vertex degrees: $L = D-A$. Note that in the context of the electrical network, an undirected, unweighted graph is considered to have resistors of $r_{ij}=1$ Ohm on each edge. However, this can be extended to weighted networks, where resistors have $r_{ij} = 1/w_{ij}$, where $w_{ij}$ denotes the edge weight between vertices $i$ and $j$. In this case, the weighted Laplacian matrix can easily be defined, and used to extend the following results.

Through applications of Kirchhoff's current and circuit laws, one can show that the effective resistance $R_{st}$ between vertices $s$ and $t$ may be calculated using a pseudoinverse of the Laplacian matrix of a graph:
\[R_{st} = (e_s-e_t)^T L^\dagger (e_s-e_t),\] where $e_i$ denotes the standard unit vector, i.e., a vector consisting of all zeros except a single $1$ in the $i^{th}$ position.
Given this expression for the effective resistance between two vertices, the total graph resistance can be expressed in terms of the sum of the reciprocals of the nonzero eigenvalues of the Laplacian matrix \cite{klein_resistance_1993}:
\begin{equation}
\label{RD}
RD(G) = \frac{1}{n} \sum_{j=2}^n \frac{1}{\rho_j}.
\end{equation}

As previously mentioned, this measure can be easily extended to weighted graphs, though it is not clear that a generalization for directed graphs is possible.

\subsubsection*{Kemeny's constant}

Given an ergodic Markov chain with transition matrix $P$, stationary vector $w$, and mean first passage times $m_{ij}$, one can define the following quantity:
\[\kappa_i = \sum_{j=1}^n m_{ij} w_j.\]
This can be interpreted as the expected time to reach a randomly-chosen state $j$, given that the chain starts in state $i$. Astonishingly, this quantity is independent of the initial state $i$ \cite{kemeny_finit_1960}. Thus the \emph{Kemeny's constant}, denoted as $K(P)$, can be computed as  
\begin{equation}
\label{Kemeny_constant}
K(P)= \sum_{j=1}^n m_{ij} w_j.
\end{equation}
An interpretation of this result is that the expected time to reach a randomly-selected destination state $j$ from a fixed initial state $i$ (where state $j$ is selected randomly according to the stationary distribution $w$) does not depend on the starting point $i$ \cite{doyle_kemeny_2009}.
Furthermore, since the $w_j$ sum to 1, we can write Kemeny's constant as a double-sum:
\[K(P) = \sum_{i=1}^n \sum_{j=1}^n w_i m_{ij} w_j,\]
admitting the interpretation of Kemeny's constant as the expected length of a random trip in the chain, where both initial and destination states are chosen randomly according to the stationary distribution. Therefore, Kemeny's constant is an intrinsic measure of a Markov chain. If the transition matrix $P$ has eigenvalues $\lambda_1=1, \lambda_2,...,\lambda_n$, then another way of computing $K(P)$ is \cite{levene_kemenys_2002},
\begin{equation} \label{Kemeny2}
K(P) = \sum_{j=2}^{n} \frac{1}{1-\lambda_j}.
\end{equation}

Note that this expression furnishes a computationally useful method for determining $K(P)$ in practice; $K(P)$ is computed as the trace of a given generalized inverse of the singular matrix $I-P$.

Kemeny's constant is a proxy for the global ``connectedness'' of a network, given the interpretation as the expected length of a random trip between states in the chain. Thus, networks characterised by small values of Kemeny's constant should be more efficient in terms of flow \cite{crisostomi_google-like_2011}. In a contact network, small values of Kemeny's constant for the random walk on the graph indicate that the individuals in the network are well-connected, while large values of Kemeny's constant may be indicative of clustered behaviour, and that it is difficult to traverse the graph. If one can identify the vertex whose removal causes the largest increase in the value of Kemeny's constant, this could be interpreted as a ``central'' vertex.

Since Kemeny's constant can be computed for any ergodic Markov chain, it extends most usefully to weighted, directed graphs.

\subsubsection*{Subdominant Eigenvalues}

For an ergodic Markov chain with transition matrix $P$, the eigenvalues may be listed $\lambda_1=1, \lambda_2, \ldots, \lambda_n$, and by Perron-Frobenius theory, $|\lambda_j|\leq1$ for all $j=2, \ldots, n$. As is evident from our discussion of Kemeny's constant above, much information regarding the dynamics of the Markov chain may be extracted from these eigenvalues. We are frequently interested in the eigenvalue $\lambda_j$ for which $|\lambda_j|$ has second-largest modulus (SLEM) after 1. Without loss of generality, suppose this is $\lambda_2$. We outline here several ways in which the value of $\lambda_2$ can be used to infer how well-connected the states of a Markov chain are (or the vertices of a graph; where the chain in question represents a random walk on a graph). In general, if $\lambda_2$ is bounded away from 1, the states of the Markov chain are considered to be ``well-connected''.

For a Markov chain with transition matrix $P$, the probability distribution after $k$ time-steps is given by $u_k^T = u_0^TP^k$, where $u_0$ is the initial probability distribution vector. It is well-known that if the Markov chain is ergodic (i.e. the matrix $P$ is primitive), then $u_k^T$ converges to the stationary distribution vector $w^T$ of the chain as $k\to\infty$, independently of the initial distribution. The asymptotic rate of this convergence is clearly dictated by the moduli of the eigenvalues of $P$. If all eigenvalues have modulus bounded away from 1, convergence happens quickly.
This convergence is often referred to as the ``mixing rate'' of the chain, and is framed in terms of how quickly the initial information is lost.

 In the context of a random walk on an undirected connected graph, there is the following bound on the total variation distance after $k$ time-steps have passed (see \cite{chung_spectral_1997}):
\[\|u_0^T P^k - w\| \leq \max_{j\neq 1} |\lambda_j|^k \frac{\max_i \sqrt{\deg(i)}}{\min_j \sqrt{\deg(j)}}.\]
As $P$ has real eigenvalues, issues arise when $\lambda_j$ is close to 1 or close to $-1$. We note that there is some difference in the dynamics of the Markov chain in these cases---a subdominant eigenvalue close to 1 in value is indicative of clustering behaviour, or of near-decoupling \cite{crisostomi_google-like_2011, hartfiel_structure_1998, breen_clustering_2018}, while $-1$ occurs as an eigenvalue of $P$ if the undirected graph is bipartite, causing periodic behaviour in the Markov chain. For this reason, in the context of disease spread in a network, we are more interested in a subdominant eigenvalue whose value is close to 1 (not whose modulus is close to 1). See \cite{breen_clustering_2018} for more discussion.

For a connected graph $G$, the transition matrix for the random walk on $G$ is $P=D^{-1}A$; denote its eigenvalues by $1\geq \lambda_2\geq \cdots \geq \lambda_n$. The subdominant eigenvalue $\lambda_2$ and its relationship to graph structure is well-studied in the context of the normalized Laplacian matrix, defined as 
\[\mathcal{L} = D^{-1/2}(D-A)D^{-1/2}\]
when $G$ has no isolated vertices. It is easily seen that $\mathcal{L}$ is similar to the matrix $I-D^{-1}A$, so that the normalized Laplacian eigenvalues $0\leq \mu_1 \leq \cdots \leq \mu_{n-1}$ are in one-to-one correspondence with the eigenvalues of $D^{-1}A$, where $\mu_j = 1-\lambda_{j+1}$. In particular, the eigenvalue $\mu_1 = 1-\lambda_2$ is often referred to as the \emph{algebraic connectivity} of the graph \cite{chung_spectral_1997} (not to be confused with Fiedler's algebraic connectivity, defined in the next section as an eigenvalue of the combinatorial Laplacian $L=D-A$). 

The eigenvalue $\mu_1$ (or the spectral gap $1-\lambda_2$) is well-known to be related to isoperimetric numbers of the graph. In particular, we have the following relationship with Cheeger's constant $h(G)$ \cite{chung_spectral_1997}:
\[\frac{h(G)^2}{2} \leq \mu_1 \leq 2h(G).\]
The Cheeger's constant is a measure of how much a vertex set can expand in the graph; formally,
\[h(G) = \min_{\substack{S\subset V(G)\\\vol(S) \leq \frac12\vol(G)}} \frac{e(S, \overline{S})}{\vol(S)},\]
where $\vol(S)$ is twice the number of edges between vertices in $S$, and $e(S, \overline{S})$ denotes the number of edges in the graph between vertices in $S$ and vertices outside $S$. We note that graphs with low values of Cheeger's constant typically have ``bottleneck'' vertices, or clusters/communities with very few edges between them. Finally, the quantity $1/ \mu_1$ is referred to as the \emph{relaxation time} of the random walk, and is studied in itself as a measure of the asymptotic rate of convergence to the stationary distribution (see \cite{aldous-fill-2014}). Note that the relaxation time $1/(1-\lambda_2)$ is the first and largest term in the expression \eqref{Kemeny2} of Kemeny's constant.

In the context of disease spread in a graph $G$, the value of the second-largest eigenvalue $\lambda_2$ of the transition matrix for the random walk (and how close it is to 1) may be used to indicate how quickly the disease may disperse through the graph, as a measure of clustering behaviour in the graph, or as a stand-in for Cheeger's constant, indicating whether the graph has good expansion properties (which would imply that the disease spreads quickly).

Since subdominant eigenvalues can be calculated for any Markov chain, this measure is easily extended to weighted and directed networks. We remark that in the case that a subdominant eigenvalue is a complex number, some work has been done in that area to show how clustering behaviour may be derived from the value of $\lambda_2$ in that case \cite{breen_clustering_2018}.

\subsubsection*{Algebraic Connectivity (AC)}

The eigenvalues of the Laplacian matrix of a graph $G$, $L=D-A$, can be listed 
$\rho_0 = 0 \leq \rho_1 \leq \cdots \leq \rho_{n-1}$, and the second-smallest eigenvalue $\rho_1$ is referred to as the \emph{algebraic connectivity} of the graph, and denoted $a(G)$ \cite{fiedler_laplacian_1989}.
This eigenvalue is strictly greater than $0$ if and only if $\mathcal{G}$ is a connected graph. The magnitude of this value reflects how well connected the overall graph is. It has been used in analysing the robustness and synchronizability of networks. There are many papers and results relating the value of $a(G)$ to many other structural quantities in the graph, such as vertex- and edge-connectivity, diameter, minimum degree, isoperimetric constants, and many others. For an overview, see the excellent survey by De Abreu \cite{de_abreu_old_2007}.
There is some research in the area of extending the definition of algebraic connectivity to directed graphs \cite{wu_algebraic_2005}. 

\subsubsection*{The Basic Reproduction Number ($\mathcal{R}_0$)}
The basic reproduction number $\mathcal{R}_0$ is perhaps the most popular indicator in the epidemiology community, and it may be defined as the expected number of individuals that a randomly infected individual can infect during his/her infection period in a fully-healthy susceptible population \cite{mei_dynamics_2017}. This value depends on the specific disease (e.g., its intrinsic infectivity), and on the topology of the network of contacts. In particular, it can be shown that in a network-SIS model (a susceptible-infected-susceptible model) $\mathcal{R}_0$ is proportional to $\lambda_{max}$, where $\lambda_{max}$ denotes the dominant eigenvalue of the adjacency matrix, and it is equal to its spectral radius \cite{mei_dynamics_2017}. Accordingly, in this paper we shall use the dominant value of the adjacency matrix as a proxy for $\mathcal{R}_0$, with the ultimate goal of testing individuals whose removal gives rise to the lowest value of $\mathcal{R}_0$

\subsection*{Modularity}
The last measure we discuss here is the \emph{modularity} of a network, which is considered to quantify the degree of community structure of the network. In order to define this measure (as defined in \cite{newman_finding_2004}), we assume there exist some pre-determined communities to which the vertices of a graph $G$ belong, and partition the vertex set accordingly; say $V(G) = S_1 \cup S_2 \cup \cdots S_r$ . We define $e_{ij}$ to be the fraction of all edges in the network that join vertices from $S_i$ with vertices from $S_j$, with $e_{ii}$ considered as the fraction of edges within the community $S_i$ (i.e. $\vol(S_i)/\vol(G)$). Define $a_i = \sum_{j=1}^r e_{ij}$, considered as the fraction of edges that connect to vertices in $S_i$. Then the modularity is defined
\[Q = \sum_{i=1}^r e_{ii} - a_i^2.\]
This quantifies the degree of community structure in the network by comparing the fraction of all edges that are within a community to the fraction of the edges connecting that community to the other communities, and observing that if there was no community structure, one could expect that $e_{ij} = a_ia_j$, giving a modularity of $Q=0$. 

In this present manuscript we do not consider modularity as one of our criticality measures used to rank nodes by the effect of their removal on this quantity. Instead, modularity plays a more vital role in our simulations, as we construct contact networks with varying levels of community structure as described by modularity. We follow the method outlined in \cite{salathe_dynamics_2010} to construct these; see section \emph{Case study: Creation of the network of daily contacts}.

\section*{Comparison for the COVID-19 case study}
\label{Comparison}

\subsection*{Case study: Creation of the network of daily contacts} \label{networkcreation}

Our case study is created according to the procedure outlined in \cite{salathe_dynamics_2010}, and later reused also in \cite{yilmaz_kemeny-based_2020}. The procedure to create the network can then be summarized in the following steps:
\begin{enumerate}
\item
6 small-world communities of 40 nodes are first created using the Watts-Strogatz algorithm \cite{watts_collective_1998}, so that each node has exactly $N_d^0$ edges connecting to nodes of the same community;
\item
We then add $30 \cdot N_d^0$ edges in a random way to connect different communities; after this step, the average node degree becomes $N_d = \frac{5}{4} N_d^0$;
\item
We then rewire \textit{between-communities} edges (i.e., edges that connect nodes belonging to two different communities) so that they become \textit{within-community} edges (i.e., edges that connect nodes belonging to the same community). In doing this, the modularity of the graph increases, and we stop the procedure once a desired level of modularity $M$ is achieved.
\end{enumerate}
In particular, we compare the different indicators for different values of $N_d$, that correspond to the number of daily contacts of individuals, and for different values of modularity $M$, that allows one to evaluate what happens for societies with milder or stronger community structures. Thus, every day we create a new graph according to the procedure previously given, and we assume that it corresponds to a network of daily contacts.

\subsection*{Case study: simulation of epidemic spread on a graph}

The previously described procedure had been proposed to mimic the network of contacts of individuals in a society, with the final objective of evaluating the impact of different testing/vaccination campaign, to better mitigate the spreading of epidemics. While such networks could be only guessed in that context, contact tracing applications are now giving the unprecedented advantage of indeed knowing when individuals meet at a close distance for a sufficient time to infect a new individual (e.g., at least 15 minutes at a distance of 1 meter). An implicit assumption here is that individuals allow the sharing of the information of their daily contacts to some centralized data center that aggregates this kind of information and knows the network (e.g., in terms of an adjacency matrix).

We then model the spreading of the virus, and the testing procedure in the following simplified way: On the first day, we randomly label two individuals as \textquotedblleft infected\textquotedblright, and they correspond to our initial condition. Every single day, we assume that a susceptible individual who gets in contact with an infectious individual, has a probability of 10 \% of being infected. Also, every day, we test $N_{test}$ individuals according to the different indicators introduced in the section \emph{Review of the state-of-the-art}: if a tested individual is found infected, then he/she is quarantined for the following 14 days, after which we assume the individual is fully recovered and not susceptible anymore (i.e., they can not be infected again). Also all his/her contacts of the same day are further tested (but not those of the previous days). During the quarantine, we assume that the individual is fully compliant with the rules, and does not have contact with any other individual. It may also happen that infectious individuals are never tested, and obviously may infect other individuals as well as they are not quarantined, and we assume that they also recover after 14 days, after which they are not susceptible anymore.

The previous model is obviously a simplification of the COVID-19 dynamics, and may be seen as a simplified agent-based SIQR model, where each individual in the population may belong to only one of the disjoint compartments of Susceptible -- Infected -- Quarantined -- and Recovered individuals. Since we only test individuals on the basis of the topology of the network, depending on the specific indicator of interest, this method may be applied in practice to target-testing individuals in a population of asymptomatic individuals, and should be obviously complemented with the good practices of testing individuals who, independently from the topology of the network, exhibit COVID-19 symptoms.

\subsection*{Discussion of the simulation results}
\begin{figure*}
\centering
\begin{tabular}[c]{l}
\includegraphics[width = 0.85\textwidth]{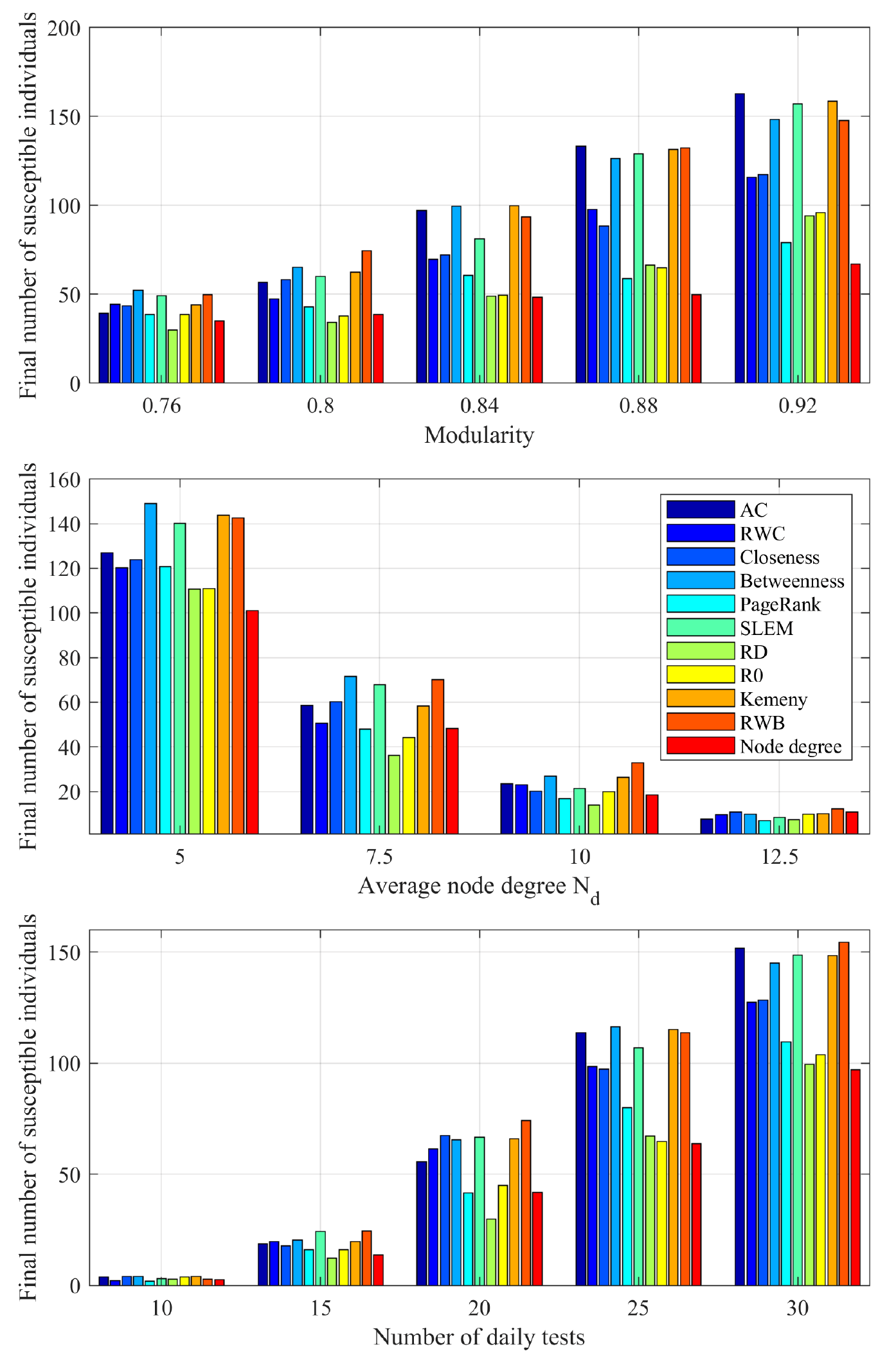}
\end{tabular}
\caption{Final number of susceptible individuals at the end of the simulation, for networks of different modularities (above), different average node degree (middle), and for different values of daily tests.}
\label{Summary_results}
\end{figure*}

\begin{figure*}
\includegraphics[width = 1\textwidth]{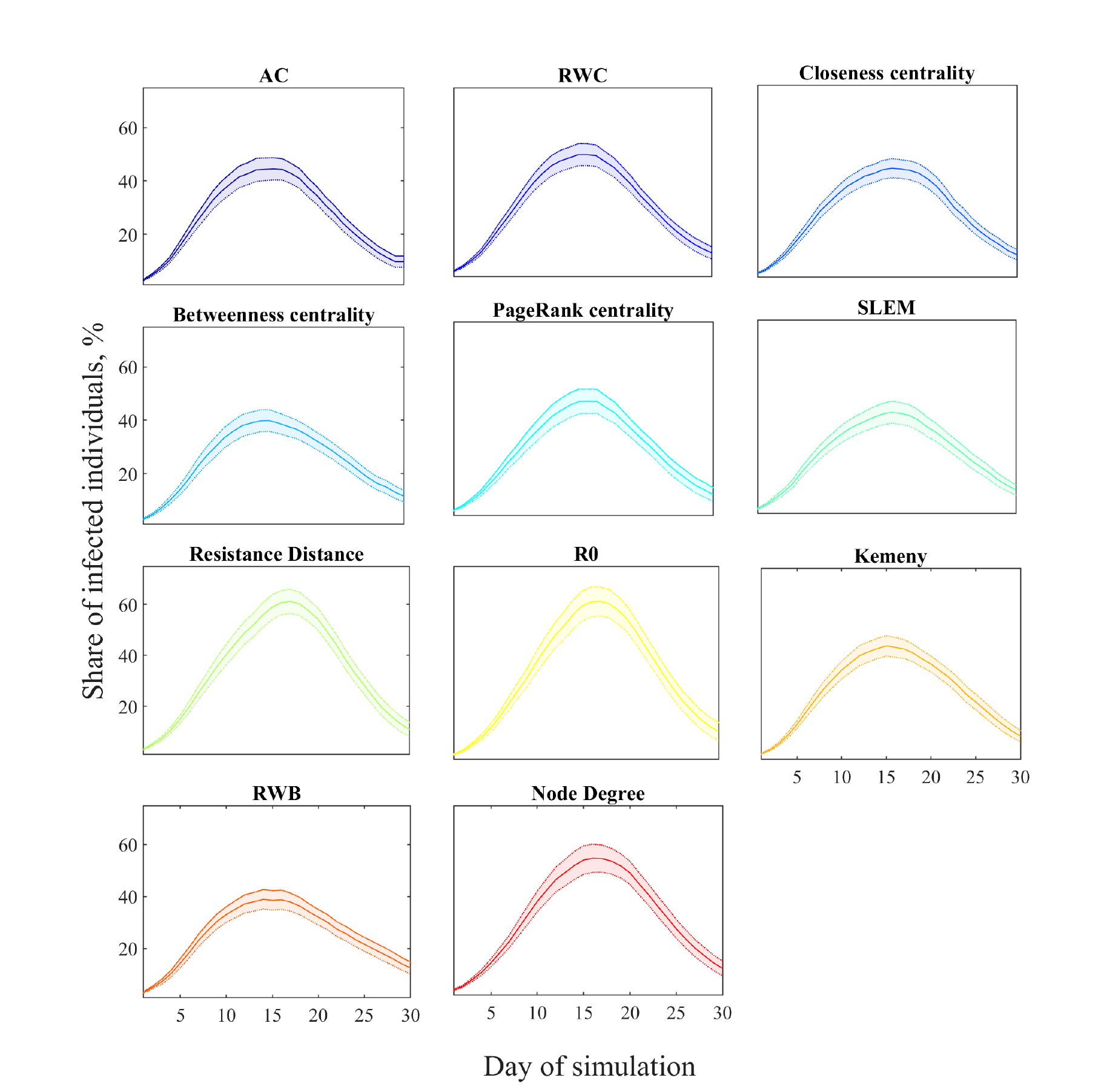}\\
\caption{Daily number of infected individuals for different indicators. The shaded area corresponds to 95\% confidence interval.}
\label{Shaded_results}
\end{figure*}

Simulation results are described in Fig. \ref{Summary_results}, where the final number of susceptible individuals at the end of the simulation (i.e., after 30 days) is shown. Note that due to the adopted COVID-19 spreading model, the final number of susceptible individuals corresponds to 240 (size of the considered population) minus the number of infected individuals (so a larger number of susceptibles individuals corresponds to a smaller number of infected individuals). Due to the stochasticity of the considered model, results shown in Fig. \ref{Summary_results} are averaged over 100 simulations, and every time two randomly chosen individuals are supposed to be infectious. The evolution of the number of infected individuals throughout the 30 days of simulations is shown in Fig. \ref{Shaded_results}.

\epstopdfDeclareGraphicsRule{.tif}{png}{.png}{convert #1 \OutputFile}
\AppendGraphicsExtensions{.tif}

In the simulations we have considered networks of modularity equal to 0.8, average node degree $N_d$ equal to 7.5 and the possibility of testing 20 different individuals every day. In particular, Fig. \ref{Summary_results} shows how performances of different ranking algorithms change if one of the previous quantities is slightly changed. In particular, from the figure above it is possible to appreciate that better results can be obtained if networks with stronger community structure are considered (values of modularity greater than 0.8). However, as from the second panel of Fig. \ref{Summary_results}, it is possible to appreciate that the most relevant quantity is the average node degree $N_d$. This could be expected since it corresponds to the number of daily relevant contacts. Finally, the last panel shows the obvious result that as the number of daily tests is increased, then a reduced number of individuals gets infected. 

\subsection*{Computational burden for different indicators}

With the increase of the network size, the computational burden could become too high for some indicators, and thus limit their scalability. Table \ref{tab:compTime} reports the computation time for the different indicators with a personal computer, equipped with a 6-core i7-8700 CPU @ 3.20GHz, RAM 16 GB. The values correspond to an average time for ten runs of the whole 30-day simulation and are sorted accordingly.

The indicators that can be computed fastest are those measuring the node centrality, i.e., Page rank, Closeness centrality, Node degree, Betweenness centrality, and RWC. As expected, node connectivity measures are more resource-demanding and result at least one order of magnitude slower. Among them, the most computationally expensive are Second Largest Eigenvalues in Modulus (SLEM) and Kemeny constant. Finally, it is noteworthy that although RWB still measures the node centrality, it is the most burdensome measure among all indicators (almost 300 times slower than Page rank). 

\begin{table}[!ht]
\centering
\begin{tabular}{|p{2cm}|p{2cm}|p{3cm}|}
\hline 
Indicator & Average computation time in seconds for 30 days & Computation method \tabularnewline
\hline 
\hline 
PageRank centrality & 3.3 & Matlab function (centrality, `pagerank')\tabularnewline
\hline 
Closeness centrality & 3.5 & Matlab function (centrality, `closeness')\tabularnewline
\hline
Node degree & 3.6 & Matlab function (degree)\tabularnewline
\hline 
Betweenness centrality & 3.8 & Matlab function (centrality, `betweenness')\tabularnewline
\hline 
Random walk centrality (RWC) & 8.3 & Based on generalized inverse (Theorem 1.1 in (a) \cite{kirkland_random_2016}) \tabularnewline
\hline 
Algebraic connectivity (AC) & 102.5 & Second smallest eigenvalue of Laplacian matrix\tabularnewline
\hline 
R0 & 111.2 & Highest eigenvalue of adjacency matrix\tabularnewline
\hline 
Resistance distance RD & 154.8 & Based on eigenvalues of Laplacian matrix - Equation (\ref{RD}) \tabularnewline
\hline 
Kemeny's constant & 179.5 & Based on generalized inverse \tabularnewline
\hline 
SLEM & 194.8 & Second largest eigenvalue modulus of a Markov chain transition matrix\tabularnewline
\hline 
Random walk betwenness (RWB) & 944.0 & Equation (\ref{RWalkBetwenness}) (the complete algorithm to compute it is described in \cite{newman_measure_2005}, section 2.2)\tabularnewline
\hline 
\end{tabular}
\caption{Computational burden to compute each indicator. Also, for replicability purposes, we explain which equations, or which Matlab function, has been used to compute it.}
\label{tab:compTime}
\end{table}

\section*{Case study: simulation on weighted graphs}
\label{Weighted_Graphs}

The primary contribution of this article is the survey and comparison of various node ranking procedures in the control of disease spread in a contact network. However, throughout the review in Section II of existing centrality indicators, we have stressed the importance of considering metrics which generalize to weighted or asymmetric (i.e., directed) graphs. This is motivated in no small part by the nature of the current pandemic and the emphasis on increased risk of transmission caused by a lengthier interaction, or a contact event with no personal protective equipment. In this section, we present a small example to motivate further consideration of this issue.

\begin{figure}
\centering
\includegraphics[width = 0.85\linewidth]{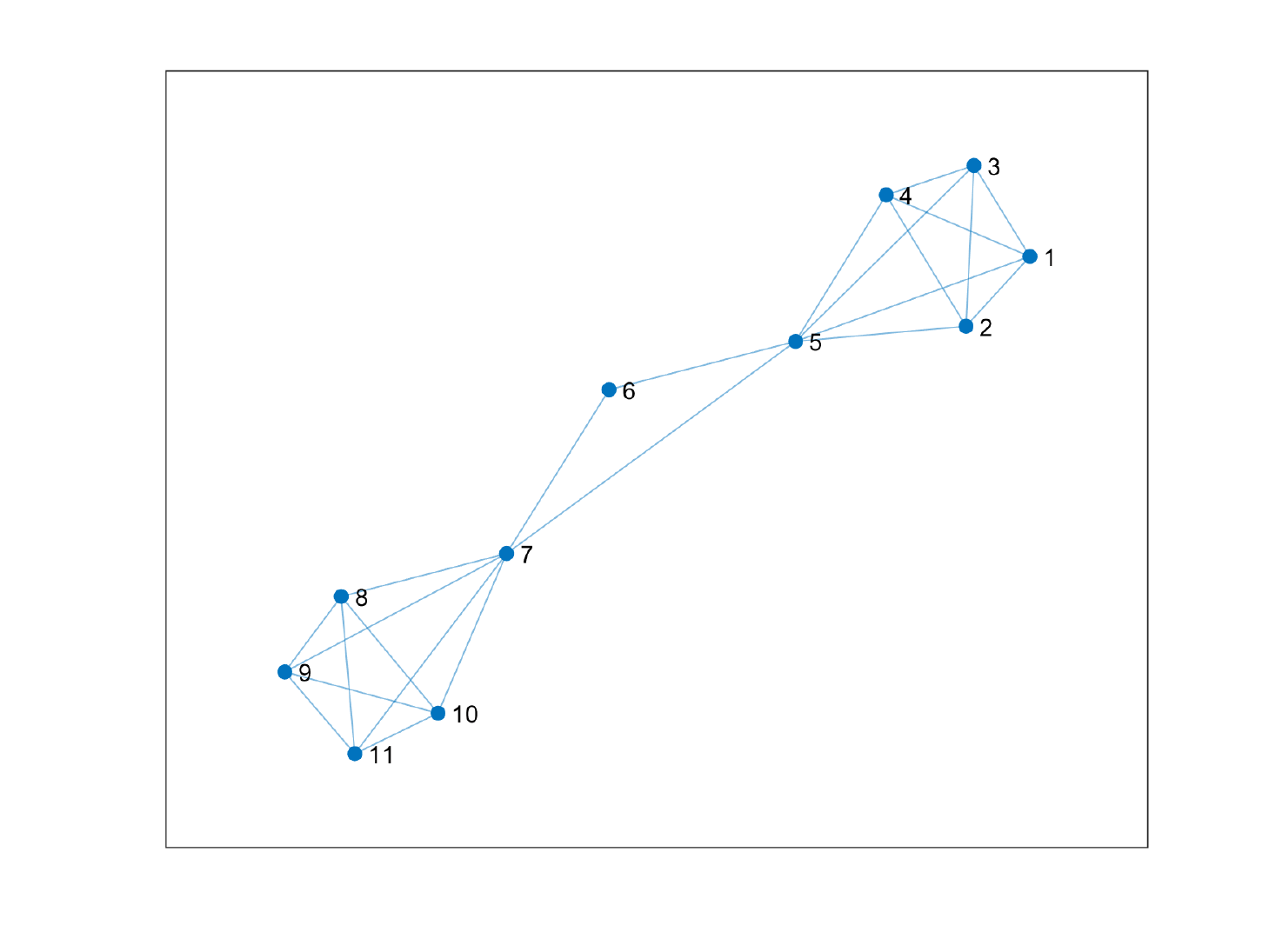}
\caption{Newman's classic example to indicate the shortcomings of shortest-path centrality measures; node 6 receives the lowest centrality ranking.}
\label{newmanrwbexample}
\end{figure}

In Newman's 2005 paper (see \cite{newman_measure_2005}), the idea of random walk betweenness was introduced, but first motivated by the example shown in Fig. \ref{newmanrwbexample}. This example indicates that while a node may not sit on any shortest path, intuition would still indicate that it is still somehow more ``central'' than others. Finding existing measures of centrality (which depended largely on shortest paths) falling short in appropriately categorizing such nodes, Newman introduced the idea of random walk betweenness. Inspired by this key example, and its natural appeal to the reader's intuition, we present a comparable example here which highlights the importance of considering the weights of connections between nodes; first by an intuitive argument, then backed up by computations.

\begin{figure}
\centering
\includegraphics[width = 0.85\linewidth]{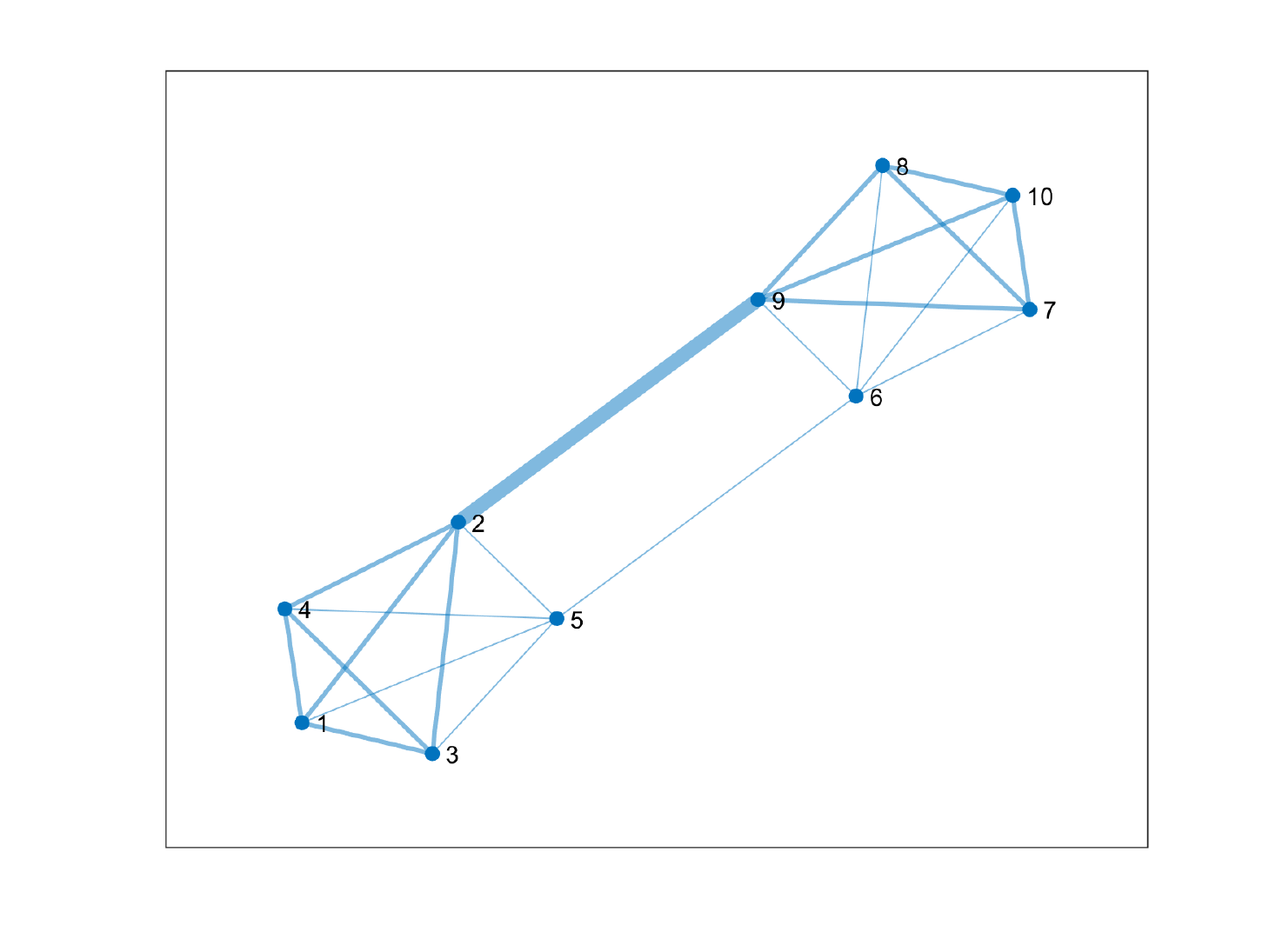}
\caption{A weighted network on 10 vertices, where weights correspond to varying probabilities of infection between individuals.}
\label{ourexample}
\end{figure}

In Fig. \ref{ourexample}, we have a network consisting of 10 nodes, and edges of different weights. The weights are given by the following matrix:
\[W = \begin{bmatrix} 0 & 0.3 & 0.3 & 0.3 & 0.1 & 0 & 0 & 0 & 0 & 0 \\
 0.3 & 0 & 0.3 & 0.3 & 0.1 & 0 & 0 & 0 & 0.95 & 0 \\
  0.3 & 0.3 & 0 & 0.3 & 0.1 & 0 & 0 & 0 & 0 & 0 \\
   0.3 & 0.3 & 0.3 & 0 & 0.1 & 0 & 0 & 0 & 0 & 0 \\
   0.1 & 0.1 & 0.1 & 0.1 & 0 & 0.1 & 0 & 0 & 0 & 0 \\
   0 & 0 & 0 & 0 & 0.1 & 0 & 0.1 & 0.1 & 0.1 & 0.1 \\
   0 & 0 & 0 & 0 & 0 & 0.1 & 0 & 0.3 & 0.3 & 0.3 \\
   0 & 0 & 0 & 0 & 0 & 0.1 & 0.3 & 0 & 0.3 & 0.3 \\
   0 & 0.95 & 0 & 0 & 0 & 0.1 & 0.3 & 0.3 & 0 & 0.3 \\
   0 & 0 & 0 & 0 & 0 & 0.1 & 0.3 & 0.3 & 0.3 & 0 
 \end{bmatrix}.\]
For our purposes, we will interpret the edge weight $w_{ij}$ as the probability of transmission from individual $i$ to $j$ (or vice versa) based on their contact time and conditions, if one of them is already infected. Note that in Fig. \ref{ourexample}, the weights are represented proportionally by thinner or thicker lines. It is a reasonable argument that based on these infection probabilities, nodes 2 and 9 should play a larger role in the spread of the disease through the entire community than nodes 5 and 6. However, if we were to assume that the graph is unweighted, or that all infection probabilities are equal, then we would assume that all four nodes (2, 5, 6, and 9) have equal importance. If testing/quarantine strategies are based on this assumption, we may fail to prevent or slow the spread of the disease.

For the remainder of this section, we analyse this example in a few ways. First, we compare the rankings of the nodes of this example according to each centrality indicator from the previous section, where we consider the network to be unweighted (i.e. all weights of contacts/edges are either 0 or 1). Then we will compute rankings when the weights of the edges are taken into account. To do so, we will need to elaborate further on how these will be computed in the weighted case. Finally, we run some simulations to compare the effectiveness of each indicator (both unweighted and weighted) in controlling the disease. These simulations differ from those in the previous section since this network is too small for those same simulations to be sensible.

\begin{figure*}

% \[  P = \begin{bmatrix}
% 0.487 &    0.154   & 0.154  &  0.154 &    0.051 &       0   &      0  &       0   &      0      &   0\\
%  0.154 &    0  & 0.154   & 0.154 &   0.051 &        0  &       0    &     0    &0.487     &    0\\
%     0.154   & 0.154    &0.487  &  0.154   & 0.051        & 0   &      0   &      0   &      0      &   0\\
%     0.154&    0.154   & 0.154    &0.487  &  0.051      &   0      &   0      &   0     &    0        & 0\\
% 0.051  &  0.051  &  0.051  & 0.051 &    0.744  &  0.051 &        0     &    0    &     0       &  0\\
%          0    &     0         &0 &        0    0.051 &   0.744 &    0.051  &  0.051  &  0.051  & 0.051\\
%          0     &    0        & 0  &       0         &0  &  0.051 &    0.487 &    0.154   & 0.154  &  0.154\\
%          0       &  0      &   0    &     0        & 0   & 0.051 &   0.154 &    0.487  & 0.154   & 0.154\\
%          0   & 0.487  &       0   &      0    &     0  &  0.051 &    0.154 &    0.154 &        0    &0.154\\
%          0     &    0   &      0       &  0     &    0   & 0.051 & 0.154&    0.154   & 0.154    &0.487
% \end{bmatrix}\]

\includegraphics[width = 0.85\linewidth]{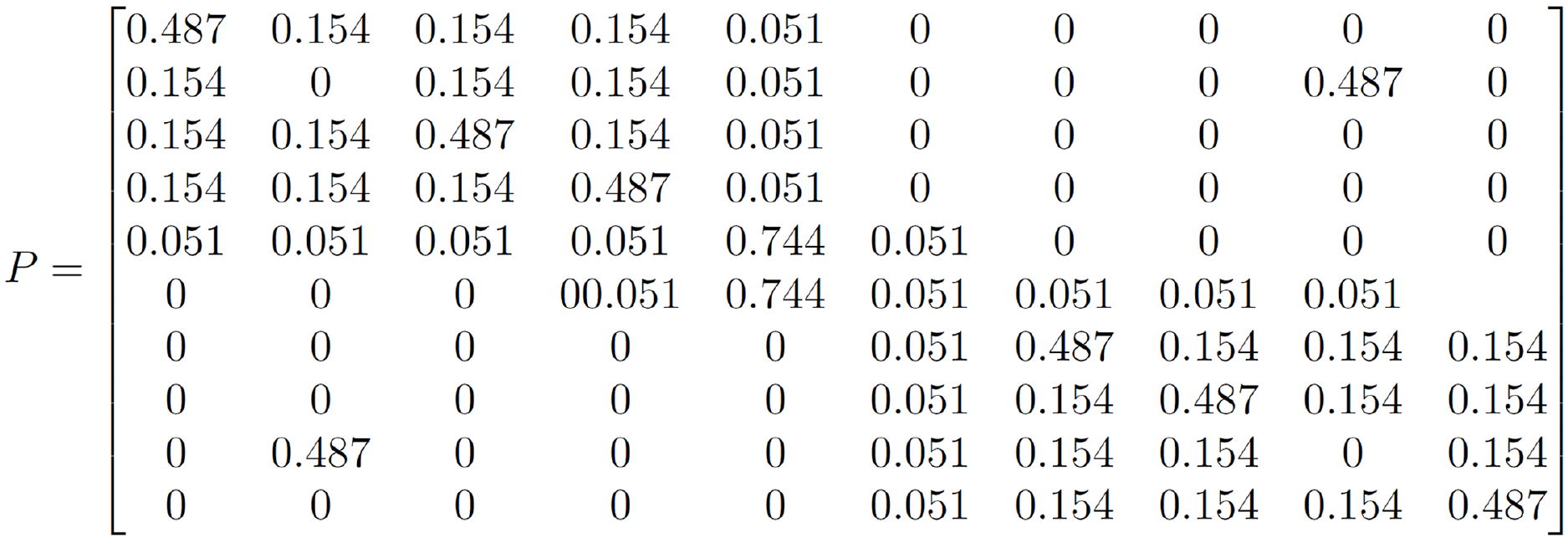}
\caption{The transition matrix for the adjusted random walk on the weighted network given in Fig. \ref{ourexample}.}
\label{lazy_matrix}
\end{figure*}

For unweighted indicators (node degree, betweenness centrality, random walk betweenness, second-largest eigenvalue, and Kemeny's constant), we compute as before, taking the weight of every edge to be 1, and working with either the $0-1$ adjacency matrix $A$, or the simple random walk on the graph with transition matrix $D^{-1}A$. For weighted indicators, we make a few adjustments:
\begin{itemize}
\item For weighted node degree, we use $W$ as the weighted adjacency matrix, and determine the weight of node $i$ as the sum of the weights of incident edges, or the sum of row $i$. 
\item In computing betweenness centrality of a node in a weighted graph, typically the edge-weights correspond to a ``cost'', and minimum distance corresponds to paths or walks of minimal cost. As such, it does not make sense in this context to compute betweenness centrality using infection probabilities as weights, since if the value of $w_{ij}$ is high, it is \emph{easier} for the disease to spread, not harder (which we would associate with high cost). We choose to simply replace the weight $w_{ij}$ of the edge between $i$ and $j$ by the cost $1-w_{ij}$, and then compute betweenness centrality for each node in the usual way.
\item There is no known formulation of random walk betweenness for weighted graphs, so we do not include a weighted version here.
\item For the second-largest eigenvalue of a probability transition matrix, and for Kemeny's constant, there are a few choices we could make for how to represent a random walk on the weighted network. One option is to simply normalize the rows of the matrix $W$, producing a stochastic matrix. However, in the disease-spread context, this may not be appropriate. Instead, to account for individuals with less-risky contacts overall, we implement an adjusted random walk, in which for certain nodes, there is a nontrivial probability that the random walker stays in place in the next time-step (in our application, this would correspond to assuming that an infectious individual will not infect another individual). To do this, we find the maximum row-sum of $W$, and normalize the entire matrix by that value. Then we replace the diagonal entry in each row by the difference between 1 and the $i^{th}$ row-sum to produce a stochastic matrix. See Fig. \ref{lazy_matrix} for the stochastic matrix representing the adjusted random walk for this example with these weights. The advantage of this normalization is that all contacts having the same duration are eventually associated with the same probability of spreading the virus in the transition matrix, while without the diagonal correction this is not guaranteed to occur, due to the normalization steps required to make transition matrices row-stochastic.
%In this example, we have the following matrix (with each entry rounded to 2 decimal places).
With this transition matrix in hand, we compute the weighted versions of these indicators using this representation of the Markov chain. For the sake of comparison, we also compute Kemeny's constant for the weighted random walk (without the diagonal correction).
\end{itemize}

All ranking indicators group the nodes of this network into at most four groups: $\{2, 9\}$, $\{5, 6\}$ and $\{1, 3, 4, 7, 8, 10\}$, where nodes in the same group receive the same rank.  The results are as follows:
\begin{itemize}
\item All unweighted indicators group $\{2, 5, 6, 9\}$ together as being equally the most critical nodes, while the remaining nodes rank second. 
\item Both weighted node degree and the weighted version of Kemeny's constant for the adjusted random walk rank $\{2, 9\}$ as most critical, $\{1, 3, 4, 7, 8, 10\}$ next, and $\{5, 6\}$ as least critical.
\item The weighted betweenness centrality measure ranks $\{2, 9\}$ as the most central nodes in the weighted network, and weights all other nodes equally.
\item The second-largest eigenvalue measure for the adjusted random walk ranks $\{2, 9\}$ as most critical, then $\{5, 6\}$, then the remaining nodes $\{1, 3, 4, 7, 8, 10\}$. Interestingly, Kemeny's constant for the weighted random walk without the diagonal correction ranks the nodes the same way.
\end{itemize}

Intuition and visual inspection of the graph are sufficient to determine that unweighted indicators fail to properly rank nodes because nodes $\{2, 9\}$ are more critical than $\{5, 6\}$ (the virus is spread more likely from one community to the other through the $\{2, 9\}$-link as the duration of their contact is longer than that of $\{5, 6\}$). Indeed, all indicators based on the weighted graph reach a consensus on indicating nodes $\{2, 9\}$ as the most critical. Conversely, there is a discrepancy on which one should be the second most critical set of nodes (i.e., Kemeny's constant indicates $\{1, 3, 4, 7, 8, 10\}$ , while the weighted betweenness centrality measure indicates $\{5, 6\}$). 

Accordingly, we run disease-spread simulations with this network so to establish which one is the correct response. As mentioned already, this network is too small to obtain meaningful results from the heuristic outlined in the previous section. Instead, we do the following: For each node, remove it from the network, infect one other node chosen at random, and run an infection simulation according to the probabilities given in $W$. Take note of the number of days it takes for the disease to spread to all the remaining 9 nodes in the network. Run this same simulation 1000 times and record the average number of days for the disease to spread to the entire network. Thus for each node in the network, we have a simulated measure of how ``critical'' it is in the spread of disease through the network - the higher the number of days for the disease to spread with node $i$ removed, the more critical node $i$ must be to the spread in the underlying network shown in Fig. \ref{ourexample}. The average numbers of days are shown for each node as follows:

\begin{center}
\begin{tabular}{cc}\hline
Node & Avg. days \\\hline
1 & 7.55 \\
2 & 17.41 \\
3 & 7.39 \\
4& 7.39 \\
5& 6.84 \\
6& 6.83 \\
7& 7.63 \\
8& 7.63 \\
9& 17.37 \\
10& 7.39
\end{tabular}
\end{center}
Note that in 1000 simulations on the full weighted network, the average number of days for the disease to spread to all ten nodes is 7.86 days. Removing some nodes might either speed up or slow down the spreading of the disease.

\subsection*{Discussion of the Results for the Weighted Case}
If we assume that the averages of the 1000 stochastic simulations of the epidemic spreading in the network provide the ``correct'' ranking of the nodes, then it is evident that indicators that were very convenient in the unweighted case, most notably those based on betweenness centrality, fail to correctly rank the nodes in the weighted cases. If we further consider that extensions of other indicators to the weighted case are not straightforward (e.g., random walk betweenness), it appears that when weighted graphs may be reconstructed (i.e., when a contact tracing app is able to store relevant information such as the durations of contacts, or the distances at which contacts occur), indicators based on the adjusted random walks such as the Kemeny's constant appear to be the most convenient choice.

\section*{Conclusions}
\label{Conclusions}
Inspired by the recent problems arising in the context of the COVID-19 pandemic, and most notably in terms of who should be tested, and who should be vaccinated first, this manuscript reviews some of the most popular ranking methodologies to identify the importance of nodes in networks of individuals. While the dynamics of the COVID-19 have been simplified, still it is possible to observe that significantly different results may be obtained if different ranking methods are adopted to select the most suitable individuals for tests or vaccination. In particular, while the actual effectiveness one strategy depends on a number of variables (most notably, the modularity exhibited by a network of individuals; by the average number of daily contacts; and by the number of available tests), it is possible to appreciate for some combinations of such variables, indicators like the algebraic connectivity, betweenness centrality, second largest eigenvalue modulus, Kemeny's constant and random walk betweenness, may actually be twice as effective than other indicators in abating the number of infected individuals (i.e., with respect to PageRank or node degree).\\
% \newline
The comparison becomes even more interesting under the assumption that durations of contacts may be measured and shared, when weighted graphs can be considered. Indeed, not all the aforementioned indicators can be generalized to work in this context. In addition, indicators that appeared to perform very well in the unweighted case, most notably, betweenness centrality and the second largest eigenvalue, fail to correctly rank nodes in the proposed simple network, which is a weighted revisitation of the classic Newman's network. Other indicators that explicitly take into account the (weighted) transition matrix, as the Kemeny's constant) appear to be the most suitable in correctly ranking the nodes.\\
% \newline
As mentioned in the introductory section, the problem of ranking nodes according to their ``importance'' arises in a number of different contexts, and different solutions have been proposed, and have become common practices, in different communities. Motivated by the recent concerns of COVID-19, this manuscript attempted at fairly comparing some of the most popular methodologies in a single application, with the ultimate objective that this comparison may inspire more work and more discussion in the field of graph theory, and ultimately provide a valuable support for testing and vaccination policies.

\section*{Acknowledgments}
J.M. acknowledges support of the OP RDE funded project CZ.02.1.01/0.0/0.0/16\_019/0000765 ``Research Center for Informatics''. S.Y. acknowledges support by TUBITAK (The Scientific and Technological Research Council of Turkey). S.K. acknowledges research support by NSERC Discovery Grant RGPIN-2019-05408. J.B. acknowledges research support by NSERC Discovery Grant RGPIN-2021-03775. M.B., P.F., R.M.-S., T.P., R.S. \& S.S. acknowledge support from EPSRC project EP/V018450/1. M.B. and T.P. also acknowledge funding support from the European Union's Horizon 2020 Research and Innovation
Programme under Grant Agreement No 739551 (KIOS CoE).

\nolinenumbers

\end{document}